\documentclass[12pt]{article}
\usepackage{amsbsy,amsmath}
\usepackage{amsfonts}
\usepackage{amssymb}
\usepackage{amscd}
\usepackage{bbm}
\usepackage{fancybox}
\usepackage{cite}
\usepackage{amsmath,amsfonts,amsbsy}
\usepackage{pstricks,pst-node}
\usepackage[small,bf,hang]{caption2}
\usepackage{graphicx}
\usepackage{epsfig}
\usepackage{psfrag}
\usepackage{comment}

\usepackage{float}

\psset{unit=1.3cm,linewidth=.5pt,radius=.2}  

\usepackage{multirow}                     
\usepackage{float}                          
\usepackage{lscape}                         
\usepackage{bm}

\newcommand{\bb}{\bar\beta}

\newcommand{\beq}{\begin{equation}}
\newcommand{\eeq}{\end{equation}}
\newcommand{\bi}{\begin{itemize}}
\newcommand{\ei}{\end{itemize}}
\newcommand{\bt}{\begin{tabular}}
\newcommand{\et}{\end{tabular}}
\newcommand{\bc}{\begin{center}}
\newcommand{\ec}{\end{center}}

\newcommand{\be}{\begin{equation}}
\newcommand{\ee}{\end{equation}}
\newcommand{\bea}{\begin{eqnarray}}
\newcommand{\eea}{\end{eqnarray}}
\newcommand{\ba}{\begin{array}}
\newcommand{\ea}{\end{array}}

\def\bbox{{\,\lower0.9pt\vbox{\hrule \hbox{\vrule height 0.2 cm
\hskip 0.2 cm \vrule height 0.2 cm}\hrule}\,}}
\newcommand{\dsl}{\pa \kern-0.5em /}

\font\mybb=msbm10 at 12pt
\def\bb#1{\hbox{\mybb#1}}

\def\bR {\bb{R}}

\def\eq#1{(\ref{#1})}

\makeatletter \@addtoreset{equation}{section} \makeatother

\def\slashchar#1{\setbox0=\hbox{$#1$}           
   \dimen0=\wd0                                 
   \setbox1=\hbox{/} \dimen1=\wd1               
   \ifdim\dimen0>\dimen1                        
      \rlap{\hbox to \dimen0{\hfil/\hfil}}      
      #1                                        
   \else                                        
      \rlap{\hbox to \dimen1{\hfil$#1$\hfil}}   
      /                                         
   \fi}

\def\eq#1{(\ref{#1})}

\begin{document}

\begin{titlepage}

\begin{center}

\vskip 1.5cm

{\Large \bf Nonlinear Electrodynamics without Birefringence}

\vskip 1cm

{\bf Jorge G.~Russo\,${}^{a,b}$ and  Paul K.~Townsend\,${}^c$} \\

\vskip 25pt

{\em $^a$  \hskip -.1truecm
\em Instituci\'o Catalana de Recerca i Estudis Avan\c{c}ats (ICREA),\\
Pg. Lluis Companys, 23, 08010 Barcelona, Spain.
 \vskip 5pt }

\vskip .4truecm

{\em $^b$  \hskip -.1truecm
\em Departament de F\' \i sica Cu\' antica i Astrof\'\i sica and Institut de Ci\`encies del Cosmos,\\ 
Universitat de Barcelona, Mart\'i Franqu\`es, 1, 08028
Barcelona, Spain.
 \vskip 5pt }
 
 \vskip .4truecm

{\em $^c$ \hskip -.1truecm
\em  Department of Applied Mathematics and Theoretical Physics,\\ Centre for Mathematical Sciences, University of Cambridge,\\
Wilberforce Road, Cambridge, CB3 0WA, U.K.\vskip 5pt }

\hskip 1cm

\noindent {\it e-mail:}  {\texttt jorge.russo@icrea.cat, pkt10@cam.ac.uk}

\end{center}

\vskip 0.5cm
\begin{center} {\bf ABSTRACT}\\[3ex]
\end{center}

All solutions of the no-birefringence conditions for nonlinear electrodynamics are found. In addition to the known Born-Infeld and Plebanski cases, we find a ``reverse Born-Infeld'' case, which has a limit to Plebanski, and an ``extreme-Born-Infeld'' case, which arises as 
a Lagrangian constraint. Only Born-Infeld has a weak-field limit, and only Born-Infeld
and extreme-Born-Infeld avoid superluminal propagation in constant electromagnetic backgrounds, but all cases have a conformal strong-field limit that coincides with the strong-field limit of Born-Infeld found by Bialynicki-Birula.

\vfill

\end{titlepage}
\tableofcontents

\section{Introduction}

The term ``nonlinear electrodynamics'' (NLED) is generally taken to mean the class of  field theories
defined by a Lagrangian density that is some Lorentz scalar function of the 2-form field-strength $F= dA$ for an abelian 1-form 
gauge potential on a four-dimensional Minkowski spacetime; no derivatives of $F$ are permitted since this would  
lead to additional (and unphysical)  propagating modes. For Minkowski coordinates  $x^\mu= (t, {\bf x})$,  all Lorentz scalars constructed from $F$ can be expressed as functions of the 
two (pseudo)scalar  quadratic Lorentz invariants\footnote{Our metric signature is
``mostly-plus''.} 
\be
\begin{aligned} 
S &= -\frac14 F_{\mu\nu}F^{\mu\nu}  = \frac12\left(|{\bf E}|^2-|{\bf B}|^2\right)\ , \\
P &= -\frac14 F_{\mu\nu}\tilde F^{\mu\nu} = {\bf E}\cdot {\bf B}\, , 
\end{aligned}
\ee
where $\tilde F$ is the Hodge dual of $F$ and $({\bf E},{\bf B})$ are its electric and magnetic components. 
The generic NLED Lagrangian density can therefore be written as $\mathcal{L}(S,P)$. Maxwell electrodynamics is recovered by choosing $\mathcal{L}=S$. 

One solution of the field equations of any NLED is a constant uniform electromagnetic field strength. If $\mathcal{L}$ is expanded about this `background' solution, the term quadratic in the perturbation $f$ of the background field strength $F$ is
\begin{eqnarray}
\mathcal{L}^{(2)} &=& {\cal L}_S\Big\{ - \frac14 f^{\mu\nu} f_{\mu\nu}\\
&& + \, \frac18 \ell_{SS} (F^{\mu\nu}f_{\mu\nu})^2 +\frac14 \ell_{SP}(F^{\mu\nu}f_{\mu\nu})(\tilde F^{\mu\nu} f_{\mu\nu} )
+ \frac18 \ell_{PP} (\tilde F^{\mu\nu} f_{\mu\nu})^2 \Big\}\, ,   \nonumber 
\end{eqnarray}
where
\be
\ell_{SS} = \frac{\mathcal{L}_{SS}}{\mathcal{L}_S} \, , \qquad \ell_{SP} =   \frac{\mathcal{L}_{SP}}{\mathcal{L}_S} \, , \qquad 
\ell_{PP} =  \frac{\mathcal{L}_{PP}}{\mathcal{L}_S} \, . 
\ee
We may assume that $\mathcal{L}_S\ne0$ because otherwise the linearized field equation does not propagate two independent polarisation wave modes; this
linear field equation is
\begin{equation}\label{linfe}
 \partial_\mu f^{\mu\nu} - \frac12 A^{\mu\nu\rho\sigma} \partial_\mu f_{\rho\sigma} =0\, , 
 \end{equation}
 where
\be
A^{\mu\nu\rho\sigma} = (\ell_{SS} F^{\mu\nu} + \ell_{SP} \tilde F^{\mu\nu})F^{\rho\sigma} + (\ell_{SP}F^{\mu\nu} + \ell_{PP} \tilde F^{\mu\nu}) \tilde F^{\rho\sigma} \, .  
 \ee
Here we are following \cite{Bialynicki-Birula:1984daz}, where it was observed that a constant uniform electromagnetic background can be  interpreted as an anisotropic optical medium provided that there is a frame in which the background Poynting vector is zero; this is the rest-frame of the medium. A typical property of anisotropic optical media is a polarisation-dependent dispersion relation for electromagnetic plane waves. This phenomenon, known as birefringence, is also a typical property of wave solutions of the equation \eqref{linfe},  but there is no birefringence in a few exceptional cases: 
\begin{itemize}

\item 
Born-Infeld electrodynamics, for which 
\be \label{L-BI}
\mathcal{L}_{\rm BI} = T - \sqrt{T^2-2TS -P^2}\, , 
\ee
where $T$ is a constant with dimensions of energy density (and an interpretation as 3-brane tension in string-theory applications). Maxwell electrodynamics is recovered in the $T\to\infty$ limit; i.e. the weak field limit.  Like Maxwell's equations, the (source-free) BI field equations have an $SO(2)$ electromagnetic duality invariance. Reality of $\mathcal{L}_{\rm BI}$ for all ${\bf B}$ requires 
\be\label{BIineq}
|{\bf E}|^2<T\, , 
\ee 
which was part of the original motivation for the theory \cite{Born:1934gh}.

\item ``Plebanski'' electrodynamics, for which
\be\label{PlebL}
\mathcal{L}_{\rm Pl} = \frac{\kappa S}{P}\, , 
\ee
where $\kappa$ is another constant with dimensions of energy density (but one that does not appear in the Lagrangian field equations). There is no weak-field limit as both ${\bf E}$ and ${\bf B}$ must be nowhere zero 
(to avoid $P=0$) but we show here that there is a strong-field limit.

\item ``Reverse Born-Infeld'' electrodynamics:
\be\label{rBI}
\mathcal{L}_{\rm rBI} = \alpha \sqrt{P^2 + 2TS -T^2} + \beta P + {\rm const}. 
\ee
for dimensionless constants $(\alpha,\beta)$. Reality for all ${\bf B}$ now requires $|{\bf E}|^2>T$ but a more useful inequality, which is both necessary and sufficient for reality, is 
\be\label{rBIineq}
T-2S < P^2/T\, . 
\ee
Despite the similarity to BI, the rBI field equations are {\it not} duality invariant (a fact that becomes manifest in the Hamiltonian formulation). 
Another difference to BI is the absence of a weak-field ($T\to\infty$) limit; there is a strong-field $(T\to0)$ limit with $\mathcal{L} \propto |P|$, but this has no wave solutions. 

For the choice of dimensionless constants $\alpha=\beta= \kappa/T$, 
where $\kappa$ is another parameter with dimensions of energy density, 
the rBI Lagrangian density is 
\be\label{rBI2}
\mathcal{L}_{\rm rBI} = \frac{\kappa }{T}\left[\sqrt{P^2+ 2TS -T^2} -P\right] \, .
\ee
The field equations are independent of $\kappa$, so $T$ remains the only natural scale for energy density. However, if we assume that $P$ is everywhere positive then we can take the $T\to0$ limit, for fixed $\kappa$, to recover \eqref{PlebL} (which provides an {\it a posteriori} justification for the assumption of positive $P$). Thus, rBI electrodynamics can be 
viewed as a generalisation of Plebanski electrodynamics. 
\end{itemize}
The first two of these zero-birefringence cases were initially found by Boillat \cite{Boillat:1966eyw,Boillat:1970gw} and Plebanski \cite{Plebanski:1970zz} from studies of shock waves, but the zero-birefringence conditions on $\mathcal{L}(S,P)$ also follow from a study of wave solutions in constant uniform electromagnetic backgrounds \cite{Bialynicki-Birula:1984daz}. The third (rBI) 
case was also found by Plebanski but in a particular Lorentz frame, which may explain why it is rarely (if ever) mentioned in citations to his 1970 lecture notes\footnote{Eq. (10.37a) of 
\cite{Plebanski:1970zz} coincides with \eqref{rBI2} in a frame for which 
$|{\bf E}\times{\bf B}|=0$ if one sets Plebanski's constants $(\tilde\mu, \tilde\nu)$ to $(\kappa,T/(2\kappa))$. His eq. (10.37b) is the ``Plebanski'' case.}. Both Boillat and Plebanski showed how the general solution to the zero-birefringence equations may be found, and they also showed that Born-Infeld is the unique zero-birefringence electrodynamics with a weak-field limit\footnote{This can also be established by straightforward power-series solution of the no-birefringence condition \cite{Mkrtchyan:2022ulc}.
Born-Infeld is also the unique duality invariant electrodynamics with ``good propagation'' properties \cite{Deser:1998wv}.}.


Here we follow the method of Boillat, which involves an interchange of the dependent and independent variables of the zero-birefringence 
conditions, leading to what we shall call the ``Boillat equations'' \cite{Boillat:1970gw}. The general solution to these equations, which is easily found, depends on integration constants, and all choices that allow a weak-field limit lead to Born-Infeld. However, there are other solutions. In particular, allowing the constant $T$ to be zero leads to the ``Plebanski'' case.  What we call ``reverse-Born-Infeld'' arises as a $T\ne0$ subcase in which a sign choice that leads to Born-Infeld is reversed. 

At the boundary between Born-Infeld and reverse-Born-Infeld, there is one other case that was not noticed by either Boillat or Plebanski. It corresponds to a solution of the ``Boillat equations'' that leads to a quadratic constraint on the Lorentz invariants $(S,P)$, rather than to a Lagrangian density. This `extremal' case leads to what we shall call:
\begin{itemize}

\item ``Extreme Born Infeld'' (eBI) electrodynamics. The Lagrangian constraint is $P^2 + 2TS -T^2=0$.
If this constraint is imposed by a Lagrange multiplier, one finds a Lagrangian density that 
can be interpreted as a non-conformal scaling limit of either BI or rBI. Because this limit takes us outside the NLED framework initially assumed, we discuss this case separately. 

\end{itemize}

We have not yet mentioned the strong-field  ($T\to0$) limit of BI electrodynamics. This is because it cannot be taken in \eqref{L-BI}. As we shall  show later, it can be taken in a Lagrangian {\sl reformulation} that includes auxiliary fields, but it was first found in the
context of the Hamiltonian formulation of BI electrodynamics, where the strong-field limit 
leads to the conformal Bialynicki-Birula (BB) electrodynamics, with an enhanced 
$Sl(2;\bb{R})$ electromagnetic duality and other unusual properties \cite{Bialynicki-Birula:1984daz,Bialynicki-Birula:1992rcm}. This example suggests that  strong-field limits of 
Plebanski and reverse-BI electrodynamics might similarly become possible in their Hamiltonian formulation, and that extreme-BI electrodynamics might have a standard Hamiltonian formulation. 
These suggestions are all realized! 

For any NLED for which $\mathcal{L}$ is a convex function of ${\bf E}$, the Hamiltonian density is defined by the Legendre transform 
\be\label{LegT}
\mathcal{H}({\bf D}, {\bf B}) = \sup_{{\bf E}} \{ {\bf E} \cdot {\bf D} - \mathcal{L}({\bf E},{\bf B})\}\, , 
\ee
where ${\bf D}$ is the (electric displacement) 3-vector field conjugate to ${\bf E}$. By construction, this Hamiltonian density is a convex function of ${\bf D}$. The convexity condition on $\mathcal{L}$ is satisfied by 
$\mathcal{L}_{\rm BI}$, and the corresponding Hamiltonian density is
\be\label{BIHamiltonian}
\mathcal{H}_{\rm BI} = \sqrt{T^2 + T(|{\bf D}|^2 + |{\bf B}|^2) + |{\bf D}\times {\bf B}|^2} -T\, . 
\ee
Ultimately, convexity is important because non-convexity implies the existence of a negative birefringence index for small amplitude waves in constant magnetic backgrounds \cite{Bandos:2021rqy}, which in turn 
implies a violation of causality for any NLED with a weak-field limit, as we show here by generalising a result of \cite{Bialynicki-Birula:1984daz} for Born-Infeld. However, exceptions may occur for NLEDs that do not have a weak-field limit, so we do not wish to impose convexity
{\it ab initio}. 

In principle, NLEDs could be defined in terms of a Hamiltonian density $\mathcal{H}$, which then gives us a first-order (phase-space) Lagrangian density 
\be\label{first-order}
\widetilde{\mathcal{L}} = {\bf D} \cdot {\bf E} 
- \mathcal{H}({\bf D},{\bf B})\, . 
\ee
Variation of ${\bf D}$ typically yields an algebraic equation for ${\bf D}$.
If this equation has a unique solution then the 
Hamiltonian field equations will be equivalent to the Euler-Lagrange equations of the Lagrangian density that we find by back-substitution. A unique solution is guaranteed if $\mathcal{H}$ is a (strictly) convex function of ${\bf D}$, and then $\mathcal{L}$ is guaranteed to be a convex function of ${\bf E}$. However, the ${\bf D}$-field equation may have a unique solution for ${\bf D}$
even if $\mathcal{H}$ is not a convex function of ${\bf D}$, in which case
elimination of ${\bf D}$ from \eqref{first-order} will still yield an equivalent second-order Lagrangian density. In this way, one can find an equivalent Hamiltonian formulation for some NLEDs whose Lagrangian density is not a convex function of ${\bf E}$. 

An example is Plebanski electrodynamics. The Hessian matrix of $\mathcal{L}_{\rm Pl}$ has eigenvalues
\be
\frac{\kappa}{P} \, , \qquad \frac{\kappa |{\bf B}|^2}{P^3} \left[S+ \sqrt{S^2+P^2}\right] \, , 
\qquad \frac{\kappa |{\bf B}|^2}{P^3}\left[S- \sqrt{S^2+P^2}\right] \, , 
\ee
and at least one is negative for any non-zero ${\bf B}$. However, there
is an equivalent first-order formulation with Hamiltonian density  
\be\label{HamPleb1}
\mathcal{H}_{\rm Pl} = \sqrt{|{\bf D}\times{\bf B}|^2 + 2\kappa ({\bf D}\cdot{\bf B}) -\kappa^2} \, .  
\ee
Provided that $\kappa\ne0$,  the ${\bf D}$-field equation of $\tilde{\mathcal{L}}$ can be solved uniquely for ${\bf D}$ in terms of $({\bf E},{\bf B})$, and the resulting (second-order) Lagrangian density is $\mathcal{L}_{\rm Pl}$. This Hamiltonian formulation of Plebanski electrodynamics still has no 
weak-field ($\kappa\to\infty$) limit but the strong-field limit ($\kappa\to0$) is now possible:
\be
\lim_{\kappa\to0} \mathcal{H}_{\rm Pl} = \mathcal{H}_{\rm BB} :=|{\bf D}\times{\bf B}| \, .  
\ee
This is the Hamiltonian density of BB electrodynamics.  Although $\mathcal{H}_{\rm Pl}$ is not a convex function of ${\bf D}$ for 
$\kappa\ne0$ (as expected since we could otherwise obtain a convex Lagrangian density by a Legendre transform) it {\it is} convex for $\kappa=0$.

Similarly, the Hamiltonian density for the particular reverse-Born-Infeld theory defined 
by \eqref{rBI2} is
\be
\mathcal{H}_{\bf rBI} = \sqrt{|{\bf D}\times {\bf B}|^2 + T|{\bf D}|^2 + 
2\kappa ({\bf D}\cdot{\bf B}) -\kappa^2} \, .  
\ee
We see directly from this result that the Plebanski case is recovered by taking the 
$T\to0$ limit for fixed $\kappa$, but it is now also possible to take the 
$\kappa\to0$ limit at fixed $T$. This yields the Hamiltonian density of 
extreme-Born-Infeld (eBI) electrodynamics:
\be\label{eBIHamiltonian}
\mathcal{H}_{\bf eBI} = \sqrt{|{\bf D}\times {\bf B}|^2 + T|{\bf D}|^2}\, . 
\ee
Equivalence to the Lagrangian formulation of eBI with a Lagrange multiplier field may be established by a Legendre transform, as we shall show later.

The above-mentioned interpretation of eBI as a non-conformal scaling limit of Born-Infeld is particularly clear in the Hamiltonian formulation. First we make the following $SL(2;\bR)$ transformation
with constant scaling parameter $\gamma$ 
\be
({\bf D}, {\bf B}) \to (\gamma^{-1}{\bf D}, \gamma\, {\bf B} ). 
\ee
The BI Hamiltonian density \eqref{BIHamiltonian} now depends on $\gamma$, but if we rescale $T\to \gamma^2T$ we can then take the $\gamma\to 0$ limit to find the eBI Hamiltonian density \eqref{eBIHamiltonian}. We could also rescale $T \to \gamma^{-2}T$ and take the $\gamma\to\infty$ limit to find the following ``magnetic'' version of eBI with 
\be\label{meBIHamiltonian}
\mathcal{H}_{\bf meBI} = \sqrt{|{\bf D}\times {\bf B}|^2 + T|{\bf B}|^2}\, . 
\ee
This NLED does {\sl not} appear as zero-birefringence case in our Lagrangian classification, but this could be because its Lagrangian formulation (which we shall give later) is `non-standard' (i.e.
not expressible purely as a function of $S$ and $P$). 

Our starting point for this route to eBI and meBI could equally well have been the reverse-BI Hamiltonian density of \eqref{rBI2}, but the route from BI correctly suggests that  the Hamiltonian densities of both eBI and meBI are convex functions of ${\bf D}$, and that the small amplitude waves propagating disturbances of constant background fields are never superluminal. These two electrodynamics theories are therefore physical, at least theoretically, and we comment in the conclusions on a possible 
application. 

The organisation of the remainder of this paper is as follows. We first review the calculation in  \cite{Bialynicki-Birula:1984daz} of the birefringence indices associated with 
the two polarisations of small-amplitude plane-wave solutions in a constant electromagnetic 
background, but with a streamlined notation and a few minor improvements. For example, the rather complicated expression for the birefringence indices presented in \cite{Bialynicki-Birula:1984daz} can be simplified significantly by noticing that its denominator factorises, 
and we extend to all NLEDs an argument of \cite{Bialynicki-Birula:1984daz} for BI that
shows why positive birefringence indices are required to avoid superluminal propagation. We also review the elegant method of Boillat for solving the zero birefringence conditions \cite{Boillat:1970gw}, relaxing restrictions that he imposed to obtain a classification of all zero-birefringence NLEDs (within the assumed Lagrangian context).  

We then move on to Hamiltonian formulations (and some alternative Lagrangian formulations). This allows a much better understanding of the relationships between the various zero-birefringence NLEDs, in addition to allowing their conformal strong-field limits to 
BB-electrodynamics.  The ``extreme-Born-Infeld'' case and its `magnetic' variant are then introduced and analysed, and their interpretation as non-conformal scaling limits of Born-Infeld is detailed. We conclude with a brief summary and discussion of our main results.

\section{Birefringence preliminaries}\label{sec:biref}

If we seek plane-wave solutions of \eqref{linfe} with wave 4-vector $k$, we find that
\begin{equation}\label{keq}
\left\{ k^\nu k_\mu + \left[\ell_{SS} G^\nu G_\mu + \ell_{SP}(\tilde G^\nu G_\mu + G^\nu \tilde G_\mu) + \ell_{PP} \tilde G^\nu \tilde G_\mu \right] \right\}\epsilon^\mu =  k^2 \epsilon^\nu\, , 
\end{equation}
where
\begin{equation}\label{Gs}
G^\mu = F^{\mu\nu}k_\nu \, , \qquad \tilde G^\mu = \tilde F^{\mu\nu} k_\nu\, . 
\end{equation}
Note the identities
\begin{equation}\label{id1}
k\cdot G \equiv 0 \, , \quad  k\cdot \tilde G \equiv 0\, ,  
\end{equation}
and 
\begin{equation}\label{id2}
G\cdot \tilde G \equiv -P k^2\, , \qquad \tilde G^2 = G^2 + 2Sk^2\, . 
\end{equation}

In the case that $k^2\ne0$, equation \eqref{keq} tells us that $\epsilon \in {\rm span}\{ k, G, \tilde G\}$ but gauge invariance allows us to choose $k\cdot\epsilon =0$, so that 
\begin{equation}\label{expandep}
\epsilon = \alpha(k) G + \beta(k) \tilde G \, . 
\end{equation}
In the case that $k^2=0$ we may expand $\epsilon$ on the basis $\{k,\tilde k,G,\tilde G\}$, where $\tilde k$ is another null 4-vector orthogonal to $G$ and $\tilde G$ but with $k\cdot\tilde k =-1$; 
now  \eqref{keq} implies that the coefficient of $\tilde k$ in the expansion of $\epsilon$ is zero, while the $k$ term is irrelevant because of gauge invariance, 
so we again have \eqref{expandep}.  Substitution of  \eqref{expandep} into (\ref{keq}) yields two equations for the two amplitudes $(\alpha,\beta)$, and after using the identities 
(\ref{id2}) to eliminate the $\tilde G$ dependence in these equations we find the matrix equation 
\begin{equation}
 \left( \begin{array}{cc} (1+ P\ell_{SP}) k^2 - \ell_{SS} G^2 & (P\ell_{SS} -2S\ell_{SP})k^2 -\ell_{SP} G^2 \\
P\ell_{PP}\, k^2 -\ell_{SP} G^2 & (1+ P\ell_{SP} -2S\ell_{PP})k^2 - \ell_{PP}G^2 \end{array}\right)\left(\begin{array}{c} \alpha \\ \beta \end{array}\right) =0\, . 
\end{equation}
The only solution is $\alpha=\beta=0$ unless the matrix has zero determinant, and this happens when 
\begin{equation}
J (k^2)^2 - 2(\Xi-S\Gamma)G^2  k^2 + \Gamma (G^2)^2 =0\, , 
\end{equation}
where
\begin{equation}
\Xi = \frac{1}{2} \left(\ell_{SS} + \ell_{PP}\right) \, , \qquad  \Gamma = \ell_{SS}\ell_{PP} - \ell_{SP}^2 \, , 
\end{equation}
and 
\be
J= 1 -P^2\Gamma + 2(P\ell_{SP}-S\ell_{PP})\, . 
\ee
This quadratic equation for $k^2$ has the solutions
\begin{equation}\label{disp}
k^2 = G^2 \lambda_\pm\, , 
\end{equation}
where the ``birefringence indices'' are
\begin{equation}\label{bir}
\lambda_\pm =  \frac{Q_\pm}{J}\, , \qquad Q_\pm = \Xi -S\Gamma  \pm \sqrt{\Psi}\, , \qquad \Psi = (\Xi -S\Gamma)^2 - \Gamma J \, . 
\ee
An equivalent formula for $\Psi$ is
\begin{equation}\label{Psi}
\Psi = (\Xi -\ell_{PP} -S\Gamma)^2 + (\ell_{SP}-P\Gamma)^2\, . 
\ee
This is the formula given in \cite{Bialynicki-Birula:1984daz}, which we have so far been following  (in a slightly different notation)  but the formula 
\eqref{bir} for the birefringence indices can be simplified, for $\Gamma\ne0$, by making use of the identity
\begin{equation}
Q_+Q_- \equiv  J\Gamma \, . 
\end{equation}
This allows us to rewrite the birefringence indices in the form 
\be
\lambda_\pm = \frac{\Gamma}{Q_\mp}\,  \qquad (\Gamma\ne0). 
\ee
For $\Gamma=0$ we have $Q_-=0$, and hence $\lambda_-=0$. This applies to conformal NLEDs \cite{Denisov:2017qou}, in particular the
unique duality invariant ModMax case \cite{Bandos:2020jsw}, since conformal invariance implies $\Gamma=0$ (but not {\it vice versa}). 

Let us now apply these results to the zero-birefringence cases of most interest here\footnote{The eBI and meBI cases must be dealt with separately as their Lagrangian formulations are non-standard.}:
\begin{itemize}

\item Born Infeld. In this case
\be 
\lambda_\pm = \lambda_{\rm BI} = \frac{1}{T-2S}\, ,    
\ee
which is positive because the BI inequality $|{\bf E}|^2<T$ implies $T-2S\ge0$. 
 
\item Reverse Born-Infeld. In this case
\be
\label{indexrbi}
\lambda_\pm = \lambda_{\rm rBI} = \frac{1}{T-2S}\, ,  
\ee
as for the BI case but the rBI inequality \eqref{rBIineq} allows $\lambda_{\rm rBI}$ to have either sign.  

\item Plebanski. In this case
\be
\lambda_\pm = \lambda_{\rm Pl} = - \frac{1}{2S}\, .  
\ee
This is the $T\to0$ limit of the rBI result, as might be expected from the
discussion following \eqref{rBI2}.  Since $S$ may be positive or negative the same is true of $\lambda_{\rm Pl}$. 

\end{itemize}

Of these cases, BI is the only one for which the birefringence index is necessarily positive.
This is significant because (as mentioned in the Introduction) positive indices are necessary for causality. 
This was demonstrated for the BI case in \cite{Bialynicki-Birula:1984daz}, but it is true more generally,
as we shall now explain. 

\subsection{Causality constraints on birefringence}

From the definition of $G$ in \eqref{Gs}, and setting $k^\mu = (\omega, {\bf k})$, we find that 
\be
G^2 = -({\bf k}\cdot {\bf E})^2 + \omega^2 |{\bf E}|^2 +2\omega {\bf k} \cdot {\bf E}\times{\bf B} + |{\bf k}\times {\bf B}|^2\, . 
\ee
Using this in the relation $k^2= G^2 \lambda$, we find the dispersion relation 
\be 
(1+ \lambda |{\bf E}|^2)\, \omega^2 + 2\lambda ({\bf k}\cdot {\bf S}) \,  \omega = (1+ \lambda |{\bf E}|^2) |{\bf k}|^2  - \lambda |{\bf k} \times {\bf E}|^2 - \lambda |{\bf k}\times {\bf B}|^2 \, ,  
\ee
where ${\bf S} = {\bf E} \times {\bf B}$.  The term linear in $\omega$ is present because the background behaves like an anisotropic optical medium in motion when 
${\bf S}$ is non-zero, so this term is absent if we choose a reference frame in which the medium is at rest.  
In this 
rest frame we may orient the coordinate axes such that
\begin{equation}\label{backEB}
{\bf E} = E\ {\bf e}_3 \, , \qquad {\bf B} =B\ {\bf e}_3\, , 
\end{equation}
for unit 3-vector ${\bf e}_3$; the dispersion relation is then
\begin{equation}\label{disp2}
\omega^2 = |{\bf k}|^2 - \frac{\lambda(E^2+B^2)}{1+\lambda E^2} |{\bf k}\times {\bf e}_3|^2 \equiv  A (k_1^2+ k_2^2) + k_3^2 \, ,
\end{equation}
where
\begin{equation}\label{defA}
A=  \left(\frac{1-\lambda B^2}{1+\lambda E^2} \right). 
\end{equation}
We might expect $A>0$ since otherwise $\omega$ is real only when $k_3^2 >|A|(k_1^2+k_2^2)$; this is certainly true in a weak-field limit since then $\lambda \approx0$, so $A\approx 1$.  For BI electrodynamics one finds that 
\be
\label{AABI}
A_{\rm BI} = \frac{T-E^2}{T+ B^2} \, , \qquad 0 < A_{\rm BI} \le 1\, , 
\ee
where the upper and lower bounds are consequences of the BI bound $|{\bf E}|^2<T$. In this case we see from \eqref{defA} that $\omega < |{\bf k}|$, so the phase-velocity is subluminal. However, it is the group velocity $d\omega/d{\bf k}$ that is relevant for causality; its magnitude is 
\begin{equation}\label{groupv}
v_g = \sqrt{\frac{A^2(k_1^2+k_2^2) + k_3^2}{A(k_1^2+k_2^2) + k_3^2}}\, , 
\end{equation}
which is less than or equal to unity for all ${\bf k}$ and all background fields iff $0<A\le1$, a condition that is satisfied for BI electrodynamics \cite{Bialynicki-Birula:1984daz}.  

The result \eqref{groupv} applies to all NLEDs, on the assumption that 
there exists a small-amplitude wave with real angular frequency $\omega$.
A sufficient condition for this is that $A>0$, in which case the group velocity will never be superluminal iff $\lambda\ge0$, which is therefore  necessary for causality.  For NLEDs with a weak-field limit the condition $A>0$ places bounds on the  electric and magnetic fields, which may be 
satisfied naturally as a consequence of the Lagrangian, as in the BI case. 

Those NLEDs without a weak-field limit, such as Plebanski and reverse-BI, must be considered separately. Using the results obtained above for their birefringence indices, we find the following results for $A$ in these two cases:
\begin{itemize}

\item Reverse Born-Infeld. As $\lambda_{\rm rBI} = \lambda_{\rm BI}$ we have
$A_{\rm rBI}= A_{\rm BI}$ but now $|{\bf E}|^2>T$, so $A_{\rm rBI}<0$. 
Wave propagation is still possible for $k_3^2 > |A_{\rm rBI}|(k_1^2+k_2^2)$, 
but it is always superluminal except when $k_1^2+k_2^2=0$ 
(in which case $v_g=1$).

\item Plebanski. As $\lambda_{\rm Pl} = -1/(2S)$, we have
\be
A_{\rm Pl} = -E^2/B^2 \le0\, . 
\ee
Wave propagation is possible only for $k_3^2 B^2 > (k_1^2+k_2^2) E^2$, and then $v_g\geq 1$, with equality holding only when $k_1^2+k_2^2=0$.
We conclude that wave propagation is generically superluminal, as might be expected  from the connection to rBI electrodynamics, and the fact that $\mathcal{L}_{\rm Pl}$ 
is not a convex function of ${\bf E}$. 

\end{itemize}

We thus conclude that Plebanski and reverse-BI electrodynamics are unphysical because (in contrast to BI electrodynamics) they allow
superluminal propagation. This conclusion leaves open the possibility of physical strong-field limits because these are not included in the class of NLEDs defined by a Lagrangian density expressible only in terms of the Lorentz invariants $(S,P)$, but this possibility is best explored via the Hamiltonian formulation of NLEDs that we consider later.

First we conclude our Lagrangian analysis with a proof that  BI, 
reverse-BI and Plebanski are the only NLEDs with Lagrangian densities $\mathcal{L}(S,P)$ for which there is no birefringence. However, we shall also find one `extremal' case as a Lagrangian constraint rather than as a Lagrangian. This is the ``extreme-BI'' electrodynamics which, as we shall see later, does {\sl not} allow superluminal propagation.

\subsection{The zero birefringence conditions}

For generic NLEDs the two birefringence indices $\lambda_\pm$ differ. The difference $\lambda_+ -\lambda_-$ is a measure of 
birefringence; i.e. the polarisation dependence of plane-wave dispersion relations. Zero birefringence occurs when $\lambda_+=\lambda_-$, and we see from
\eqref{bir} that this occurs when $Q_+=Q_-$; equivalently, when $\Psi=0$. From the expression  \eqref{Psi} for $\Psi$ we see that 
NLEDs with zero birefringence are those for which the following two equations hold: 
\be
\Xi -\ell_{PP} =S\Gamma\, , \qquad \ell_{SP} = P\Gamma\, . 
\ee
These are equivalent to the two equations
\be\label{zerobcon} 
\begin{aligned}
\mathcal{L}_S (\mathcal{L}_{SS} - \mathcal{L}_{PP}) &= 2S(\mathcal{L}_{SS}\mathcal{L}_{PP}-\mathcal{L}_{SP}^2) \\
\mathcal{L}_S \mathcal{L}_{SP} &= P(\mathcal{L}_{SS}\mathcal{L}_{PP}-\mathcal{L}_{SP}^2)\, . 
\end{aligned}
\ee
We recall that $\mathcal{L}_S$ is assumed to be non-zero. We shall also assume that $\Gamma$ is non-zero because zero birefringence for $\Gamma=0$ yields only the free-field Maxwell case. 

To solve the equations \eqref{zerobcon} we shall, following Boillat \cite{Boillat:1970gw}, convert them into equations
for which the independent variables are $(\mathcal{L}_S, \mathcal{L}_P)$ rather than $(S,P)$.  
For presentational purposes, it is convenient to make the replacements
\be
(S,P) \to (x,y) \, , \qquad (\mathcal{L}_S, \mathcal{L}_P)  \to (u,v)\, . 
\ee
The equations \eqref{zerobcon} now become 
\be\label{B-eqs}
\begin{aligned}
u (u_x - v_y) &= 2x\,\varpi \, , \\
u u_y &= y\,  \varpi\,  , 
\end{aligned}
\ee
where 
\be
\varpi = u_x v_y - u_y v_x \, . 
\ee
Notice that symmetry of mixed partial derivatives of $\mathcal{L}$ implies 
\be\label{sigchi-id} 
u_y \equiv v_x\, . 
\ee
Notice too that $\varpi$ is the Jacobian for a change of the independent variables from $(x,y)$ to $(u,v)$, so that 
\be
\left(\begin{array}{cc} x_u & x_v \\ y_u & y_v \end{array}\right) = \frac{1}{\varpi} \left(\begin{array}{cc} u_y & - v_y \\ - v_x & u_x \end{array} \right) \, . 
\ee
In particular, 
\be
\frac{u_x -v_y }{\varpi} = y_v- x_u \, , \qquad \frac{u_y}{\varpi} \equiv \frac{v_x}{\varpi} = -y_u\, ,  
\ee
which allows us to rewrite the zero birefringence conditions \eqref{B-eqs} as the following equivalent ``Boillat equations'' 
\be\label{B-Eqs}
u ( y_v - x_u) = 2x \, , \qquad u \, y_u = -y \, .  
\ee

\subsection{General solution of the Boillat equations}\label{subsec:Boillat}

The general solution of the second of the Boillat equations can be written as
\be
y = \frac{f'(v)}{u} \, \quad \Rightarrow \quad y_u = -\frac{f'(v)}{u^2} \, , \qquad y_v= \frac{f^{\prime\prime}(\chi)}{u} \, . 
\ee
for some function $f(v)$.   As $y_u\equiv  x_v$, we also have 
\be
x_v = -\frac{f'(v)}{u^2} \,  \qquad \Rightarrow \quad x=- \frac{f(v)}{u^2} + g(u) \,,
\ee
for some function $g$.  Now we have $(x,y)$ as expressions involving the two functions $f(v)$ and $g(u)$. Substituting these expressions  into the first of the Boillat equations \eqref{B-Eqs},  we find that 
\be
f^{\prime\prime}(v) =u g^\prime(u) + 2g(u)\, . 
\ee
This requires both sides to equal  the same constant, which we call $T$. Thus,
\be
f^{\prime\prime}(v) =T \, , \qquad u g^\prime(u) + 2g(u)=T \, , 
\ee
and hence 
\be
 f(v)= \frac{T}{2}v^2 + \kappa v+ c\, , \qquad g(u) = \frac{T}{2}  + \frac{c'}{u^2}\, ,
\ee
where $(\kappa,c,c')$ are the additional integration constants.  This gives us the complete solution for $(x,y)$:
\be\label{xysol} 
x =   \frac{T}{2} -\frac{\tfrac12 T v^2 + \kappa v + (c-c')}{u^2} \, , \qquad y= \frac{T v+ \kappa}{u} \, . 
\ee
There are two cases (and their subcases) to consider:
\begin{itemize}

\item $T\ne 0$. In this case the second of the equations \eqref{xysol} can be written as 
\be
Tv = u y -\kappa \, . 
\ee
 Using this in the first of equations \eqref{xysol}  we find that 
 \be\label{return}
 u^2 ( T^2 -2Tx-y^2) = 2T(c-c') -\kappa^2\, . 
 \ee
 The right hand side of this equation is an arbitrary constant, which may be 
 positive, negative or zero. We shall discuss these three subcases in turn.
 
 \begin{enumerate} 
 \item $u^2 ( T^2 -2Tx-y^2) = (Ta)^2$ for non-zero dimensionless constant $a$. In this case
 \be
 u= \frac{aT}{\sqrt{T^2-2Tx-y^2}} \, , \qquad v= \frac{ay}{\sqrt{T^2-2Tx-y^2}} - \frac{\kappa}{T} \, . 
 \ee
 These are equations for $\mathcal{L}_S$ and $\mathcal{L}_P$ that are easily integrated to give
 \be
 \mathcal{L} = -a \sqrt{T^2-2TS -P^2} - \frac{\kappa P}{T} + {\rm const}. 
 \ee
 The term linear in $P$ is a total derivative that does not contribute to the field equations. If we choose $a=1$ and the constant term such that $\mathcal{L}(0,0)=0$,  we have the Born-Infeld Lagrangian density:
  \be
 \mathcal{L}_{\rm BI}=  T - \sqrt{T^2-2TS -P^2} \, . 
 \ee
 We now assume $T>0$ to ensure convexity.

 
 \item $u^2 ( T^2 -2Tx-y^2) = - (\kappa a')^2$ for non-zero dimensionless constant $a'$. In this case
 \be
 u= \frac{a' \kappa}{\sqrt{y^2+2Tx -T^2}}\, , \qquad v= \frac{\kappa}{T} \left[ \frac{ay'}{\sqrt{y^2+2Tx -T^2}} -1\right]\, .
 \ee
 Integration now yields 
\be
\mathcal{L}  = \frac{\kappa}{T}\left[ a'\sqrt{P^2 +2TS -T^2} - P\right] + \rm const. 
\ee
This is the reverse BI Lagrangian density, parametrized by dimensionless constants $a'$ and $\kappa/T$ instead  $(\alpha,\beta)$. 
Setting $a'=1$, we recover the specific rBI case of \eqref{rBI2};
it can be shown that convexity does not hold in this case for either sign of $T$.


\item $u^2 ( T^2 -2Tx-y^2)=0$. As $u=\mathcal{L}_S$, which is assumed to be non-zero, we get a Lagrangian constraint equivalent to 
\be
T^2 -2TS -P^2=0\, . 
\ee
If this constraint is imposed with a Lagrange multiplier field then we get the Lagrangian density for what we call ``extreme Born-Infeld'' (eBI) electrodynamics. Because of the dependence on an additional scalar field (the Lagrange multiplier) this lies outside the class of NLEDs defined by Lagrangian densities $\mathcal{L}(S,P)$, although we shall see later that 
its Hamiltonian is a non-conformal scaling limit of the BI Hamiltonian. 
 
\end{enumerate}

\item $T=0$. In this case 
\be\label{zeroT}
u^2 x= - \kappa v- (c-c')\, , \qquad u y = \kappa \, . 
\ee
The second of these equations is just $P\mathcal{L}_S =\kappa$, which is trivially integrated to 
\be
\mathcal{L} = \kappa\left(\frac{S}{P}\right) + h(P) \, , 
\ee 
for some function $h$. We then have
\be
\mathcal{L}_P = -\kappa \frac{S}{P^2} + h'(P) \qquad \left(\Leftrightarrow  \quad v = - \frac{\kappa x}{y^2} + h'(y)\right) \, .  
\ee
Using these results in the first of equations \eqref{zeroT}, and assuming that $\kappa\ne0$ (since otherwise we are led to conclude that $\mathcal{L}_S=0$)  we find that $h$ is linear in $P$, and hence a constant plus a total derivative term
which we can ignore. We thus find that 
\be
\mathcal{L} = \frac{\kappa S}{P} \, , 
\ee
which is the Plebanski case\footnote{Using the complex Riemann-Silberstein
variables, a Lagrangian for Born-Infeld similar in form to  the Plebanski Lagrangian was found by Schr\" odinger
in 1935  \cite{Schrodinger:1935oqa}.
We thank Karapet Mkrtchyan for bringing this to our attention.}. 

\end{itemize}


\section{Hamiltonian and strong-field limits}

The Hamiltonian density of BI electrodynamics can be obtained by the Legendre transform, as defined in \eqref{LegT}. The result is
\be
\label{biH}
\mathcal{H}_{\rm BI} = \sqrt{T^2 + T(|{\bf D}|^2 + |{\bf B}|^2) + |{\bf D}\times{\bf B}|^2} -T\, . 
\ee
Lorentz invariance is not manifest but electromagnetic duality {\it is} manifest because this acts by a phase-rotation of 
the complex 3-vector field ${\bf D} +i{\bf B}$.  In the weak-field limit ($T\to\infty$) we recover Maxwell electrodynamics, with 
$\mathcal{H}_{\rm Max}= \tfrac12(|{\bf D}|^2 + |{\bf B}|^2)$. In the 
strong-field limit we find Bialynicki-Birula electrodynamics, with 
\be
\mathcal{H}_{\rm BB} = |{\bf D}\times{\bf B}|\, , 
\ee 
which has an $Sl(2;\bb{R})$-duality invariance. 

We now aim to find the Hamiltonian formulations of the other zero-birefringence theories of electrodynamics, and investigate the relations between them and their conformal limits. A useful check on results is the condition for a Hamiltonian density to define a Lorentz invariant theory. If we express $\mathcal{H}$ as a function of the three rotation invariants
\be
s= \frac12(|{\bf D}|^2 + |{\bf B}|^2)\, , \qquad \xi= \frac12(|{\bf D}|^2 - |{\bf B}|^2)\, , \qquad \eta= {\bf D}\cdot {\bf B}\, , 
\ee
then the field equations will be Lorentz invariant iff \cite{Bialynicki-Birula:1984daz}
\be
\mathcal{H}_s^2 - \mathcal{H}_\xi^2 - \mathcal{H}_\eta^2 =1 \, . 
\ee
Moreover, $\mathcal{H}$ will be electromagnetic duality invariant iff it can be written as a function of the two 
duality invariants $s$ and $\xi^2+ \eta^2$ (or $p^2 = s^2- \xi^2 -\eta^2 \equiv |{\bf D} \times {\bf B}|^2$). 

We shall also show in this section that the inclusion of auxiliary scalar fields allows us to find equivalent  Lagrangian formulations that allow strong-field limits to be taken in a manifestly Lorentz invariant way. In the BI case, an appropriate starting point is the following
Ro{\v c}ek-Tseytlin (RT) Lagrangian density in which two auxiliary scalar fields $(u,v)$ are introduced in order to linearize the 
dependence of the BI Lagrangian density on the Lorentz scalars $(S,P)$ \cite{Rocek:1997hi}: 
\be
\mathcal{L}_{\rm RT}= - \frac{T}{2}\left\{ v + v^{-1}(1+u^2)\right\} +vS + uP\, . 
\ee
The equations found from varying $(u,v)$ may be solved for $(u,v)$ (uniquely up to an overall sign), and back-substitution yields
\be
\mathcal{L}_{\rm RT} \to  \mp \sqrt{T^2-2TS -P^2}\, , 
\ee
which is the standard BI Lagrangian density for the upper sign choice after addition of the constant $T$. However,
we may now take the $T\to0$ limit to arrive at the Lagrangian formulation of BB electrodynamics proposed in \cite{Bialynicki-Birula:1992rcm} in which the scalar fields $(u,v)$ now impose two Lagrangian constraints:
\be\label{BBLag3}
\mathcal{L}_{\rm BB}= vS + uP\, .  
\ee
Equivalence to the Hamiltonian formulation was verified at the level of the field equations in \cite{Bialynicki-Birula:1992rcm}. It
may also be verified by a Legendre transform of the BB Hamiltonian density \cite{Bandos:2020hgy};  although the ``canonical''  Lagrangian 
density found in this way is identically zero, its domain is restricted by the Lagrangian constraints $S=0$ and $P=0$, so a solution of 
the variational problem defining the Legendre transform by the method of Lagrange multipliers yields precisely \eqref{BBLag3}. 

We shall see that similar alternative Lagrangian formulations can be found for other NLEDs discussed here, allowing the strong-field limit to be taken; in every case this yields the BB Lagrangian density of \eqref{BBLag3}.

\subsection{Plebanski electrodynamics}  

As explained in the introduction, the closest that one can get to a Hamiltonian formulation of Plebanski  electrodynamics is a first-order Lagrangian of the form 
\be\label{tildeL}
\widetilde{\mathcal L}_{Pl} =   {\bf D} \cdot {\bf E} -\mathcal{H}_{\rm Pl} \, , 
\ee
where ${\bf D}$ is an auxiliary field, and $\mathcal{H}_{\rm Pl}$ is given by \eqref{HamPleb1}; i.e.
\be\label{HamPleb1'}
\mathcal{H}_{\rm Pl} = \sqrt{|{\bf D}\times{\bf B}|^2 + 2\kappa ({\bf D}\cdot{\bf B}) -\kappa^2} \, . 
\ee
There is still no weak-field ($\kappa\to\infty$) limit. However, we can take the strong-field ($\kappa\to0$) limit to arrive at the 
Hamiltonian density of BB electrodynamics:
\be
\mathcal{H}_{\rm BB} = |{\bf D}\times{\bf B}|\, . 
\ee

An expression equivalent to \eqref{HamPleb1'} is 
\be
\mathcal{H}_{\rm Pl} =\kappa \sqrt{-{\rm det } M} \, , 
\ee
where $M$ is the $3\times 3$ symmetric matrix with entries 
\begin{equation}
    M_{ij}=\delta_{ij}- \kappa^{-1}  (D_i B_j+D_j B_i)\ . 
    \end{equation}
This matrix has eigenvalues 
\be
 \lambda_\pm= 1 -\kappa^{-1}\left[ {\bf B}\cdot {\bf D} \mp  
 |{\bf B}| |{\bf D}|\right]\ ,   \qquad \lambda_3= 1\ ,  
\ee
and we can use this to show that 
\be\label{HamPleb2}
\mathcal{H}_{\rm Pl}= \sqrt{ |{\bf B}|^2 |{\bf D}|^2 - \big( \kappa - {\bf B}\cdot {\bf D}\big)^2 }\, ,   
\ee
which is indeed equivalent to \eqref{HamPleb1'}. 

The equivalence of the first-order Lagrangian density $\widetilde{\mathcal L}_{Pl}$ 
to the original Plebanski Lagrangian density of \eqref{PlebL} may be established by elimination of ${\bf D}$ by means of its field equation, which is
\be\label{Deq}
\mathcal{H}_{\rm Pl}\,  {\bf E} =  |{\bf B}|^2 {\bf D} + \left(\kappa-{\bf D}\cdot {\bf B} \right){\bf B} \, . 
\ee
Contracting this equation with ${\bf B}$ we find that 
\be\label{PPP}
P= \frac{\kappa |{\bf B}|^2}{\mathcal{H}_{\rm Pl}} \, . 
\ee
As $|{\bf B}|^2\ne0$, we conclude that $P\ne0$ (for nonzero $\kappa$) and that its sign is fixed to be the sign of $\kappa$. Next, we take the norm squared of both sides of \eqref{Deq} to deduce that
\be
|{\bf E}|^2= \frac{|{\bf B}|^2}{\mathcal{H}_{\rm Pl}^2}  \left[ |{\bf D}\times{\bf B}|^2 + \kappa^2\right] \, , 
\ee
which shows that ${\bf E}$ cannot be zero, again as expected because the Plebanski Lagrangian density is also singular for zero ${\bf E}$. Using this result,  we obtain an expression for $S$:
\be
S= \frac{\kappa(\kappa- {\bf D}\cdot{\bf B})|{\bf B}|^2}{\mathcal{H}_{\rm Pl}^2}\, .
\ee
Combining this with \eqref{PPP} we find that 
\be
\kappa- {\bf D}\cdot{\bf B} = \frac{\kappa S|{\bf B}|^2}{P^2}\, , 
\ee
which we may use in \eqref{Deq} to deduce that 
\be
{\bf D}= \frac{\kappa}{P} \left[ {\bf E} - \left(\frac{S}{P}\right) {\bf B} \right]\, . 
\ee
This is the unique solution of the ${\bf D}$ field equation, so back-substitution into \eqref{tildeL} yields a classically equivalent Lagrangian density. It is straightforward to verify that this is the Plebanski Lagrangian density $\kappa S/P$, restricted to $P>0$ (given $\kappa>0$) because of \eqref{PPP}.

It might appear from this result that the strong-field limit of Plebanski to BB can only be taken in the Hamiltonian formulation. 
However, it is possible to introduce auxiliary fields into the Lagrangian formulation in a way that allows the strong-field limit to be 
taken in a manifestly Lorentz invariant way, as we  showed above for the BI case. 
Consider the following Lagrangian density involving a scalar field $\varphi$ and a pseudoscalar $\chi$:
\be\label{linearized}
\mathcal{L}=\varphi S +\chi \varphi P - \kappa\chi \, . 
\ee
The field-equations for $(\varphi,\chi)$ are 
\be
S=- \chi P\, , \qquad \varphi P = \kappa\, , 
\ee
which we can solve for $(\varphi,\chi)$ on the assumption that $P\ne0$ and $\kappa\ne0$. Back-substitution then yields the Plebanski Lagrangian density $\kappa S/P$.  

Notice that parity is broken only by the last term in \eqref{linearized}. As this is proportional to $\kappa$, we get a parity-invariant theory by setting 
$\kappa=0$. This yields the Lagrangian density 
\be\label{BBLag2}
\mathcal{L}_{\rm BB} = \varphi S + \varsigma P \qquad (\varsigma = \chi\varphi) \, ,  
\ee
which is the BB Lagrangian density of \eqref{BBLag3} with $(u,v)= (\varsigma,\varphi)$. 
Thus, the new Lagrangian formulation of Plebanski electrodynamics provided by \eqref{linearized} not only linearizes the dependence on the Lorentz scalars $(S,P)$ but also makes manifest the strong-field limit to BB-electrodynamics.


\subsection{Reverse Born-Infeld}

The reverse-Born-Infeld theory defined by the Lagrangian density 
$\mathcal{L}_{\rm rBI}$ of \eqref{rBI} has a first-order formulation:
\be\label{rBI'}
\widetilde{\mathcal{L}}_{\rm rBI} = {\bf D} \cdot {\bf E} - \mathcal{H}_{\rm rBI}\, , 
\ee
where  
\be
\label{HamrBI}
\mathcal{H}_{\rm rBI} = \sqrt{|{\bf D}\times{\bf B}|^2 
+ T\left[|{\bf D} - \beta {\bf B}|^2 - \alpha^2 T(T+ |{\bf B}|^2)\right]} + {\rm const}.  
\ee
Not surprisingly, the only effect of the term linear in $P$  in $\mathcal{L}_{\rm rBI}$ is to shift ${\bf D}$ by $\beta {\bf B}$.  One may see from this result that rBI electrodynamics is {\sl not} invariant under $SO(2)$ `duality' rotations of $({\bf D},{\bf B})$, in contrast to the BI case. 

By choosing the coefficients $(\alpha,\beta)$ as we did in \eqref{rBI2}
we find the Hamiltonian density corresponding to the particular rBI Lagrangian density of \eqref{rBI2}: 
\be\label{rBI2ham}
\mathcal{H}_{\rm rBI} = \sqrt{|{\bf D}\times{\bf B}|^2 + T|{\bf D}|^2
+ 2\kappa ({\bf D}\cdot{\bf B}) -\kappa^2}\, . 
\ee
Notice that the $T\to0$ limit, for fixed $\kappa$, yields the Plebanski Hamiltonian density of \eqref{HamPleb2}, in agreement with the Lagrangian analysis of this limit 
in the Introduction.

To verify equivalence of the first-order and second-order Lagrangian formulations of rBI 
we must eliminate ${\bf D}$ from \eqref{rBI'} using the ${\bf D}$ field equation. This is straightforward but can be simplified by introducing an auxiliary field; we explain this here for the particular rBI Hamiltonian density of \eqref{rBI2ham}. The first step is to notice that this Hamiltonian density is equivalent to 
\be
\mathcal{H}'_{\rm rBI}= \frac12 \left\{ e^{-1}\left[(T+ |{\bf B}|^2) |{\bf D}|^2
-  (\kappa -{\bf D}\cdot{\bf B})^2\right] + e\right\}\, ,
\ee
where $e$ is an auxiliary field. Elimination of $e$ yields $\mathcal{H}_{\rm rBI}$, 
but we may now eliminate  $e$ {\sl after} elimination of ${\bf D}$ by means of its field equation: 
\be\label{eqD}
{\bf D} = \frac{1}{(T+ |{\bf B}|^2)} \left\{ e{\bf E} - 
(\kappa - {\bf D}\cdot{\bf B}){\bf B}\right\} \, .
\ee
Contracting this with ${\bf B}$ (and subtracting $\kappa $ from both sides) we find that
\be\label{contracted}
 ({\bf D}\cdot{\bf B})- \kappa  =  \frac{eP}{T} - \frac{\kappa(T+|{\bf B}|^2)}{T}\ .
\ee

Contracting \eqref{eqD} with ${\bf E}$ yields
\be
{\bf D} \cdot {\bf E} = \frac{1}{(T+ |{\bf B}|^2)}\left[e|{\bf E}|^2 - 
(\kappa - {\bf D}\cdot{\bf B})P \right] \, , 
\ee 
and taking the norm-squared on both sides of \eqref{eqD} yields
\begin{eqnarray}
&& e^{-1} \left[(T+ |{\bf B}|^2)|{\bf D}|^2 - (\kappa - {\bf D}\cdot{\bf B})^2\right] = \nonumber 
\\
&& \frac{1}{(T+ |{\bf B}|^2)}\left[ e |{\bf E}|^2 - 2(\kappa - {\bf D}\cdot{\bf B})P -
e^{-1} T(\kappa - {\bf D}\cdot{\bf B})\right]\, . 
\end{eqnarray}
Combining these results yields
\be
\mathcal{L}_{\rm rBI} ' = \frac{1}{2(T+|{\bf B}|^2)} \left\{ e (2S-T) + 
e^{-1}T ({\bf D}\cdot{\bf B}-\kappa )^2\right\}\, , 
\ee
where $ ({\bf D}\cdot{\bf B}-\kappa)$ is  the expression given  in \eqref{contracted}; this reduces to
\be\label{predeform}
\mathcal{L}'_{\rm rBI} = \frac{\kappa}{2T} \left\{ \tilde e^{-1} [P^2+2TS -T^2] + \tilde e\right\}  - \frac{\kappa P}{T} \, , 
\ee
where
\be
\tilde e= \frac{e}{\kappa(T+|{\bf B}|^2)} \, . 
\ee
Elimination of $\tilde e$ now yields
\be
\label{rrbi}
\mathcal{L}_{\rm rBI} = 
\frac{\kappa}{T} \left[ \sqrt{P^2 + 2TS -T^2} - P\right] \, ,  
\ee
which is the rBI Lagrangian density of \eqref{rBI2}.

\section{Extreme Born-Infeld}

If we rescale the electromagnetic fields, 
\be\label{rescaling}
({\bf E},{\bf B}) \to \gamma ({\bf E},{\bf B}) \, , 
\ee
and set $T= \gamma^2\tilde T$, for some positive constant $\gamma$, we find that 
\be\label{Lto0}
\mathcal{L}_{\rm BI} \to \gamma^2 \left\{ \tilde T - \sqrt{\tilde T^2 -2\tilde TS -P^2 }\right\} \, , 
\ee
In other words, the BI Lagrangian density is rescaled as expected but with $T\to\tilde T$. The $\gamma\to0$ limit is one in which $\mathcal{L}_{\rm BI}\to0$, but we may first replace $\mathcal{L}_{\rm BI}$ by the equivalent Lagrangian density involving an auxiliary scalar field $e$:
\be
\mathcal{L}'_{\rm BI} = \gamma^2\left\{\tilde T - \frac12 \left[\frac{1}{\gamma^2 e} (\tilde T^2 -2\tilde TS -P^2) \pm \gamma^2 e\right] \right\}  \, .  
\ee
For the upper sign, elimination of $e$ returns us to \eqref{Lto0}, while the same procedure for the lower sign yields a similar rescaling of the rBI Lagrangian density for particular parameters. Irrespective of this sign choice, the $\gamma\to0$ limit now yields the Lagrangian density of ``extreme BI'' electrodynamics; after setting $e=-1/\ell$ and renaming $\tilde T$ as $T$, this is
\be\label{ebiLag}
\mathcal{L}_{\rm eBI}  = \frac12 \ell\,   (T^2 -2TS -P^2) \, . 
\ee
What was an auxiliary field is now a Lagrange multiplier for the Lagrangian constraint of the 
`extremal' case in our classification of solutions to the Boillat equations in subsection \ref{subsec:Boillat}; this was the $T\ne0$ subcase `between' BI and rBI, so it is not surprising that 
it can be found as a limit of both BI and rBI. 

As stated in subsection \ref{subsec:Boillat}, the eBI case is one that takes us beyond the 
initial definition of NLEDs in terms of Lagrangian densities that can be  written as 
a function of the Lorentz invariants $(S,P)$ {\sl only}. However, its Hamiltonian formulation is 
standard. To pass to the Hamiltonian 
formulation we define 
\be
\label{hfg}
{\bf D} :=  \frac{\partial \mathcal{L}_{\rm eBI}}{\partial {\bf E}} = \ell (T {\bf E}+P {\bf B})\ .
\ee
Contracting with ${\bf B}$, we deduce that
\be\label{solveP}
P= \frac{{\bf D}\cdot{\bf B}}{\ell(T+|{\bf B}|^2)}\ .
\ee
A contraction with ${\bf E}$ yields
\be\label{defDeBI}
{\bf D}\cdot{\bf E}= \frac{1}{\ell T (T+|{\bf B}|^2)} \left[ |{\bf D}|^2 (T+|{\bf B}|^2)- ({\bf D}\cdot{\bf B})^2\right]\ .
\ee
Finally, by taking the norm-squared of both sides of \eqref{hfg} and using \eqref{solveP} we find that
\be
T^2{\bf E}^2 =\frac{1}{\ell^2(T+|{\bf B}|^2)} \left[ |{\bf D}|^2(T+|{\bf B}|^2) 
-({\bf D}\cdot{\bf B})^2\right]\, ,  
\ee
and hence (by subtracting $T^2|{\bf B}|^2$ from both sides) an expression for $S$ in terms of $({\bf D},{\bf B})$. Using these results and 
then defining
\be
\tilde\ell :=  \ell T(T+|{\bf B}|^2)\, , 
\ee
we find that
\be
\mathcal{H}'_{\rm eBI} ={\bf D}\cdot{\bf E}-\mathcal{L}_{\rm eBI}= 
\frac12\left\{ {\tilde \ell}^{-1} \left[|{\bf D}|^2(T+|{\bf B}|^2) -({\bf D}\cdot{\bf B})^2\right] + \tilde \ell \right\}\ ,
\ee
where the prime reminds us that we have still to eliminate $\tilde \ell$ by its algebraic field equation. This last step gives the result
\be
\label{ebiHam}
\mathcal{H}_{\rm eBI}= \sqrt{|{\bf D}\times{\bf B}|^2 + T |{\bf D}|^2}\ . 
\ee
Notice that this reduces to the BB Hamiltonian density when $T=0$, so eBI could be
viewed as a non-conformal deformation of BB electrodynamics. As we shall see, BB and eBI
electrodynamics have significant features in common that are not shared with other
zero-birefringence NLEDs. 

We pause here to consider how the rescaling of \eqref{rescaling} is 
realized in the Hamiltonian formulation. The definition of ${\bf D}$ in \eqref{defDeBI} 
is preserved by the rescaling (with a rescaled $T$) if ${\bf D}\to \gamma^{-1}{\bf D}$, so 
the Lagrangian rescaling of $({\bf E},{\bf B})$ becomes the following  $Sl(2;\bb{R})$ transformation of 
$({\bf D},{\bf B})$:
\be\label{Hamrescale}
({\bf D},{\bf B}) \to (\gamma^{-1} {\bf D}, \gamma\, {\bf B})\, .  
\ee
It is obvious that this transformation reproduces $\mathcal{H}_{\rm eBI}$ 
with $T$ replaced by $\tilde T =T/\gamma^2$. 

It is also obvious that $\mathcal{H}_{\rm eBI}$ is not duality-invariant, which tells us that the limit from BI to eBI must break this symmetry. This is easily verified in the Hamiltonian formulation. After the rescaling  \eqref{Hamrescale} the BI Hamiltonian density becomes 
\be\label{scaledBI}
\sqrt{|{\bf D}\times {\bf B}|^2 + \tilde T |{\bf D}|^2 + 
\gamma^4 \left[\tilde T|{\bf B}|^2 +\tilde T^2\right]} -\gamma^2 \tilde T\, .  
\ee 
For any finite non-zero $\gamma$, this is duality invariant with rescaled $SO(2)$ duality transformations. However, the limit $\gamma\to 0$, with $T\to0$ for fixed  $\tilde T$ (which can then be renamed as $T$) breaks duality invariance and yields the eBI Hamiltonian density. Further rescaling just reproduces the eBI Hamiltonian density with a rescaled $T$, as we have already seen.

If we had chosen to view eBI as a limit of rBI rather than BI then the above discussion of how and why eBI breaks duality invariance would not have arisen, and this might suggest that the limit to eBI from rBI 
is more natural. However, there is an important feature of $\mathcal{H}_{\rm BI}$,  {\sl not} shared with $\mathcal{H}_{\rm rBI}$, that {\sl is} preserved by the limit to $\mathcal{H}_{\rm eBI}$, and that is convexity as a function of ${\bf D}$. The Hessian matrix of $\mathcal{H}_{\rm eBI}$ has eigenvalues 
\be
0\ ,\qquad \frac{T+|{\bf B}|^2}{\mathcal{H}_{\rm eBI}} \, , \qquad 
\frac{T|{\bf D}|^2(T+|{\bf B}|^2)}{(\mathcal{H}_{\rm eBI})^3} \, . 
\ee
For $T>0$, there is one zero eigenvalue but the other two are positive, implying convexity; not ``strict convexity'' because the Hessian matrix has zero determinant, and this is because $\mathcal{H}_{\rm eBI}$ is a homogeneous function of ${\bf D}$ of unit degree; i.e. 
${\bf D} \cdot {\bf E} = \mathcal{H}_{\rm eBI}$, where 
\be\label{eebi}
{\bf E} := \frac{\partial \mathcal{H}_{\rm eBI}}{\partial {\bf D}} \ .
\ee
A corollary is that the `canonical' Lagrangian density, defined by Legendre transform of 
$\mathcal{H}_{\rm eBI}$, is zero. Nevertheless, we can still recover the eBI Lagrangian density 
of \eqref{ebiLag} by a Legendre transform because \eqref{eebi} restricts the domain of the zero 
`canonical' Lagrangian by imposing a constraint on ${\bf E}$, which is 
\be\label{dBBlagcon}
P^2+2TS-T^2=0 \ .
\ee
The variational problem inherent in the definition of the Legendre transform may then be solved by the method of Lagrange multipliers, and this leads directly to the Lagrangian density $\mathcal{L}_{\rm eBI}$ of \eqref{ebiLag}, following steps spelled out for the BB case (i.e. $T=0$) in \cite{Bandos:2020hgy}. Whereas the Hessian matrix has two zero eigenvalue for $T=0$, which leads to the two Lagrangian constraints 
of \eqref{BBLag3}, now we have only one zero eigenvalue and hence only one Lagrangian constraint.

\subsection{Birefringence for eBI}

We have seen that the eBI theory is a non-conformal scaling limit of the BI theory. We can use this result to deduce the birefringence index of eBI from the BI index. 

From the definition of birefringence indices in \eqref{disp} we see that the 
scaling \eqref{rescaling}, which implies $S\to \gamma^2S$, requires
$\lambda \to \gamma^{-2}\lambda$. Taking this into account, the BI 
index equation $\lambda_{\rm BI} = 1/(T-2S)$ is reproduced
by the scaling but with $T\to \tilde T$, as one would expect. 
Taking the $\gamma\to0$ limit for fixed $\tilde T$, which we can then rename as $T$, we get the 
eBI birefringence index: 
\be\label{lameBI}
\lambda_{\rm eBI} = \frac{1}{T-2S} = \frac{T}{P^2} >0\, , 
\ee
where the second equality follows from the eBI constraint $T^2-2TS=P^2$. 

We are now able to investigate possible violations of causality via superluminal propagation in constant electromagnetic backgrounds. If \eqref{lameBI} is used to compute 
the factor $A$ appearing in the dispersion relation of \eqref{defA}, we find that 
\be
A_{\rm eBI}= \frac{E^2-T}{B^2+T}\, , 
\ee
where we have used the fact that $P^2= E^2B^2$ in the rest-frame of the background; 
recall that this was assumed in the derivation of the dispersion relation.  
However, in this frame the constraint $P^2+2TS-T^2=0$ implies $E^2=T$, and 
hence $A_{\rm eBI}=0$. This implies lightlike wave propagation, although waves can only propagate in a direction that is (anti)parallel to the
co-linear background electric and magnetic fields. Recalling the 
inequality $0< A_{\rm BI} \le1$, and the fact that $A_{\rm BI}\approx 0$ when 
$T^2 -2TS -P^2\approx 0$  (which is $E^2\approx T$ in the background rest-frame) we see that eBI is (as the name indicates) an extreme limit of BI. It 
is presumably for this reason that we find no evidence of causality violation in eBI electrodynamics.

The strong-field $(T\to 0)$ limit of eBI is BB, so one might wonder whether we could use this fact to determine the birefringence index of BB electrodynamics. Naively, we would simply take the $T\to0$ limit of $\lambda_{\rm eBI} = T/P^2$, to get $\lambda_{\bf BB} =0$, but $P=0$ is a Lagrangian constraint of the BB theory, so the limit is ill-defined.

\subsection{Magnetic eBI}

The BI Hamiltonian density of \eqref{scaledBI} with rescaled fields can also be written as 
\be\label{scaledBIm}
\sqrt{|{\bf D}\times {\bf B}|^2 + \check T |{\bf B}|^2 + 
\gamma^{-4} \left[\check T|{\bf D}|^2 +\check T^2\right]} -\gamma^{-2} \check T\, ,  
\ee 
where
\be
\check T = \gamma^2 T = \gamma^4 \tilde T\, . 
\ee
We may now take the $\gamma\to\infty$ limit, with $T\to0$ for fixed $\check T$. This yields 
a `magnetic' version of eBI; after renaming $\check T$ as $T$, the Hamiltonian density is
\be\label{BBB}
\mathcal{H}_{\rm meBI} = \sqrt{|{\bf D}\times{\bf B}|^2+T |{\bf B}|^2} \, . 
\ee
This is  also a convex function of ${\bf D}$ because the eigenvalues of the Hessian matrix are 
$$
0\, , \qquad \frac{|{\bf B}|^2}{\mathcal{H}}\, , 
\qquad \frac{T|{\bf B}|^4}{\mathcal{H}^3}\, . 
$$
The reason for the zero eigenvalue in this case is that the Hamiltonian depends only 
on two of the three components of ${\bf D}$. 

To pass to the Lagrangian formulation we first introduce the electric field: 
\be\label{nnn}
{\bf E} :=\frac{\partial \mathcal{H}_{\rm meBI} }{\partial {\bf D}}=\frac{|{\bf B}|^2 {\bf D}-({\bf D}.{\bf B}) {\bf B}}{\sqrt{|{\bf D}\times{\bf B}|^2+T |{\bf B}|^2  }} \ .
\ee
The contraction with ${\bf B}$ yields the constraint $P=0$. The `canonical' 
Lagrangian density is 
\be 
\mathcal{L}^{({\rm can})}_{\rm meBI} = {\bf D} \cdot {\bf E} - \mathcal{H}_{\rm meBI}
= - \sqrt{-2TS} \, , 
\ee
which restricts its domain to $|{\bf E}|\le |{\bf B}|$. 
Incorporating the constraint $P=0$ with a Lagrange multiplier field $\ell$, we arrive at the 
meBI Lagrangian density: 
\be\label{mdbb}
\mathcal{L}_{\rm meBI} =-\sqrt{-2TS}+\ell P\  \qquad (S<0). 
\ee

To verify that the Hamiltonian density is recovered by a Legendre transform, it is convenient to rewrite \eq{mdbb} in the form 
\be\label{mdbbb}
\mathcal{L}'_{\rm meBI} = e^{-1}S +\ell P -\frac{T}{2}e  \, , 
\ee
where $e$ is an auxiliary field; its elimination returns us to \eqref{mdbb}. 
We note, in passing, that $T\to0$ yields the BB Lagrangian density of \eqref{BBLag2}, 
as we should expect since the $T\to0$ limit of \eqref{BBB} yields the BB Hamiltonian 
density. As a check, we can take the Legendre transform of $\mathcal{L}'_{\rm meBI}$ 
to arrive at the following Hamiltonian density: 
\be
\mathcal{H}'_{\rm meBI} = \frac12 
\left\{e^{-1}|{\bf B}|^2 + e \left[ |{\bf D} -\ell {\bf B}|^2 +T\right]\right\}\, , 
\ee
where the prime reminds us that we must now eliminate the auxiliary fields $(e,\ell)$.
This yields the result of \eqref{BBB}. 

Using the same logic that allowed us to deduce $\lambda_{\rm eBI}$, we similarly deduce that 
\be
\lambda_{\rm meBI} = \lim_{\gamma\to\infty} (\gamma^{-4}\check T-2S)^{-1} = - \frac{1}{2S}\, , 
\ee
as for the Plebanski case.  However, to apply this result to wave propagation in the background rest frame (zero Poynting vector) we must set ${\bf E}$ to zero because there is no non-zero ${\bf E}$ in this frame that satisfies both $S<0$ and $P=0$. We then find that $A_{\rm meBI}=0$, and hence lightlike propagation, but only in a direction (anti)parallel to the background magnetic field, exactly as for eBI.

\section{Summary and Discussion}

The Born-Infeld (BI) theory of electrodynamics has many remarkable features.
One is its relevance to the dynamics of D3-branes of IIB superstring theory, in which context the constant with dimensions of energy density introduced by Born and Infeld can be interpreted as the D3-brane tension $T$. 
The feature of most relevance to this paper is the absence of 
birefringence in constant electromagnetic backgrounds. 
It is surprising that there is {\sl any} nonlinear electrodynamics theory (NLED) with identical dispersion relations for the two polarizations of all 
small-amplitude plane-wave disturbances of such backgrounds because the background breaks rotational symmetry. This makes it of interest to understand better the conditions that imply zero birefringence. Potentially, they are linked to other very special features of physical theories in which zero-birefringence NLEDs arise; e.g. supersymmetry in the case of the D3-brane. 

A natural question in this context is whether there are NLEDs other than Born-Infeld without birefringence. For the class of NLEDs defined by a Lagrangian density that is a function of the two Lorentz invariants $S$ (a scalar) and $P$ (a pseudoscalar) constructible from the electric and magnetic fields $({\bf E},{\bf B})$,  this question has been investigated previously, notably by Boillat and Plebanski. In particular, Boillat showed that the zero-birefringence conditions on $\mathcal{L}(S,P)$ are equivalent to a very simple pair of differential equations, here called the ``Boillat equations'', and that natural assumptions on integration constants, essentially equivalent to the requirement of a weak-field limit, leads uniquely to Born-Infeld.

All other solutions to the Boillat equations yield electrodynamics theories that, in contrast to BI, have no weak-field limit; they are also not duality invariant. One solution yields a 
zero-birefringence NLED found by Plebanski that has generally been discarded as unphysical; we agree with this assessment because we have shown that it permits superluminal propagation of small amplitude disturbances of constant electromagnetic backgrounds. Another case that we have called ``reverse-Born-Infeld'', is similarly unphysical. 

There remains one `extremal' solution to the Boillat equations that leads to a Lagrangian constraint rather than a Lagrangian. This case is best understood in the Hamiltonian formulation because a Legendre transformation of the Hamiltonian density yields a ``canonical'' Lagrangian density that is identically zero, but simultaneously imposes a constraint on the domain of this zero function; this constraint is the one arising from the `extremal' solution of the Boillat equations. Incorporating this constraint using a Lagrange multiplier field, we find a `non-standard'  Lagrangian density for what we have called ``extreme Born-Infeld'' (eBI) electrodynamics. The name derives from the fact that eBI is a non-conformal scaling limit of BI, as can be most easily seen in the Hamiltonian formulation. As this relation to BI suggests, eBI  is a physical theory with a convex Lagrangian density and no superluminal propagation in a constant electromagnetic background. We have also found a `magnetic' analog of eBI electrodynamics, which is similarly obtainable as non-conformal scaling limit of Born-Infeld, and has similarly good physical  properties.   

In general, the class of NLEDs defined by a Hamiltonian density $\mathcal{H}({\bf D},{\bf B})$, where ${\bf D}$ is conjugate to the electric field ${\bf E}$, is larger than the class of NLEDs defined by a Lagrangian density $\mathcal{L}({\bf E},{\bf B})$, even after the conditions required for $\mathcal{H}$ to define a Lorentz invariant theory have been imposed. All of the zero-birefringence NLEDs on the list just summarised have a Hamiltonian formulation, and this greatly clarifies the relationships between them. It also shows that {\sl all} NLEDs of the list have a strong-field limit to the conformal Bialynicki-Birula (BB) electrodynamics, which was first found as a strong-field limit of Born-Infeld electrodynamics.

Although the strong-field limit to BB electrodynamics is most easily seen in the Hamiltonian formulation, one can find alternative Lagrangian densities depending on additional auxiliary scalar fields for which this strong-field limit is manifest. We have presented such Lagrangian densities for several of the NLEDs considered here, including BI. A common feature is that auxiliary fields 
serve to linearize the dependence on $(S,P)$ and the strong-field limit then leaves a Lagrangian density that consists only of two Lagrange multipliers that impose the constraints $S=0$ and $P=0$, but this is a Lagrangian formulation of BB electrodynamics \cite{Bialynicki-Birula:1992rcm}. A Legendre transform of the BB Hamiltonian density yields a ``canonical'' Lagrangian density that is identically zero, but with a domain restricted by the Lagrangian constraints $S=0$ and $P=0$ \cite{Bandos:2020hgy}. The eBI case is similar, but with only one Lagrangian constraint; it is
naturally viewed as a deformation of BB electrodynamics, as well as a scaling limit of BI electrodynamics.  

Finally, let us recall that one of the principal applications of Born-Infeld electrodynamics is to the dynamics of Dp-branes in string theory, in particular the D3-brane of IIB superstring theory. The physics of open strings in a constant electric background  exhibits interesting features as the background electric field approaches its ``critical'' value: $|{\bf E}|^2\to T$, for zero magnetic field.  Open strings, which may be viewed as dipoles carrying opposite electric charges at their ends, are stretched along the direction of the electric field until, at the critical value, the `effective' string tension vanishes. This effect has been exploited in the past to construct non-commutative open-string theories \cite{Seiberg:2000ms,Gopakumar:2000na,Russo:2000zb}. We expect that the 
extreme-BI theory constructed in this paper, and perhaps its `magnetic' variant too,  is relevant to the description of open strings in this ``critical'' regime. We hope to explore this possibility in the future.

\section*{Acknowledgements}
We thank Iwo Bialnycki-Birula for helpful correspondence, and Igor Bandos, Kurt Lechner, Karapet Mkrtchyan, Dmitri Sorokin and Arkady Tseytlin for helpful comments. We are grateful to Sergei Kuzenko for sending us a photocopy of Plebanski's 1970 lecture notes on nonlinear electrodynamics. 
PKT has been partially supported by STFC consolidated grant ST/T000694/1. JGR acknowledges financial support from a MINECO
grant PID2019-105614GB-C21.



\providecommand{\href}[2]{#2}\begingroup\raggedright\endgroup


\end{document}